# Photonic Neuromorphic Computing enabled by a BIC Metasurface

Jingsong Fu[1, †], Ruiheng Jin[1, †], Zhaohui Xie[1, †], Haijun Tang[1, †, *], Xiong Jiang[1], Yue Cui[1], Xiangtong Kong[1], Wentao Hao[1], Geyang Qu[1], Can Huang[1,3,5, *], Qinghai Song[1,2,3,4,5, *]

[1] Ministry of Industry and Information Technology Key Lab of Micro-Nano Optoelectronic Information System, Guangdong Provincial Key Laboratory of Semiconductor Optoelectronic Materials and Intelligent Photonic Systems, Harbin Institute of Technology, Shenzhen 518055, China.

[2] Pengcheng Laboratory, Shenzhen 518055, China.

[3] Quantum Science Center of Guangdong-Hongkong Macao Greater Bay Area, Shenzhen 518055, China.

[4] Collaborative Innovation Center of Extreme Optics, Shanxi University, Taiyuan 030006, Shanxi, China.

[5] Heilongjiang Provincial Key Laboratory of Advanced Quantum Functional Materials and Sensor devices, Harbin Institute of Technology, Harbin 150001, China.

[†] These authors contribute equally to this work.

*Corresponding author: qinghai.song@hit.edu.cn; tanghaijun@hit.edu.cn; huangcan@hit.edu.cn.

**Abstract:**

Photonic neuromorphic computing promises revolutionary advances in parallel and high-speed processing, yet a key challenge persists: co-integrating nonlinearity, dense connectivity, and intrinsic memory monolithically to enable brain-inspired, spatiotemporal information processing. Here, we overcome this challenge by introducing a monolithic photonic recurrent network based on an active metasurface operating at bound state in the continuum (BIC). The BIC mode mediates strong, long-range coupling across the lattice, creating a reconfigurable recurrent network topology in hardware. Concurrently, the gain medium provides both optical nonlinearity for neuronal activation and a finite carrier lifetime that serves as a built-in, analog temporal memory. This synergy enables computation to emerge directly from the collective spatiotemporal dynamics of the driven-dissipative photonic system, effectively realizing a physical reservoir computer on a chip. We experimentally validate a minimal yet physically complete system on benchmark tasks—brain MRI image classification and human action recognition—achieving 92.16% and 85.36% accuracies, respectively. This work establishes a scalable pathway toward ultrafast, energy-efficient neuromorphic intelligence where processing is an inherent property of tailored light–matter interaction.

**Introduction**

The relentless expansion of artificial intelligence is driving computing beyond the deterministic, sequential paradigm of Von-Neumann architectures and toward a brain inspired paradigm that thrives on parallelism, adaptability, and the intrinsic dynamics of physical systems [1-3]. At the heart of this neuromorphic shift lies a quest not merely for faster hardware, but for substrates that can natively encode and process information in ways fundamentally aligned with biological computation—particularly in harnessing collective system dynamics for complex spatiotemporal tasks [4,5].

Among physical substrates, photonics stands out for its inherent speed and bandwidth [6-8]. However, most current photonic neuromorphic approaches remain largely confined to emulating the feedforward structure of deep neural networks, mapping static layers onto static photonic circuits [9,10]. While effective for certain classification tasks, this strategy fundamentally overlooks a deeper capability inherent to biological and advanced physical networks. Crucially, implementing such recurrent computation in hardware requires a platform that embodies two co-dependent features: a reconfigurable, high-dimensional network topology enabling rich signal mixing across space, and intrinsic, nonlinear temporal dynamics within each node to serve as an analog memory [11-13]. This presents a dual challenge for integrated photonics. First, creating reconfigurable long-range connections on-chip is difficult. While optical resonators or microlasers offer excellent nonlinear responses, their interactions are typically confined to the near field through subwavelength coupling [14-18]. External feedback loops can break locality but at the cost of integration density and system complexity [19-21]. Second, embedding usable temporal memory is nontrivial. Many photonic nonlinearities are inherently instantaneous, forcing reliance on external delays that create bottlenecks in speed and energy efficiency [22-24]. An ideal platform would therefore co-locate programmable connectivity and non-instantaneous dynamics within a monolithic all-optical medium.

Here, we demonstrate that such a photonic neuromorphic computing platform can be realized in an active BIC metasurface [25–27]. The long-range coupling inherent to BIC mode creates a reconfigurable recurrent network [26]. Simultaneously, the carrier and photon dynamics of the gain medium provide a built-in, analog temporal memory [28-30], enabling the system to function as a physical reservoir for information processing. Computation thus emerges not from programmed instructions, but from the orchestrated spatiotemporal dynamics of this driven-dissipative photonic system. As a proof of concept, we validate this physical computing paradigm on two benchmarks: brain MRI image classification (Kaggle dataset) and human action recognition in video (NTU RGB+D dataset). The system achieves accuracies of 92.16% and 85.36%, respectively. These results establish a practical and scalable route toward monolithic photonic intelligence, where computation is an intrinsic property of tailored light–matter interaction.

**Physical Implementation of the neural Network**

**Figure 1** schematically illustrates how the core computational primitives of the proposed network—nonlinear activation, reconfigurable long-range coupling, and temporal memory—are physically co-integrated in the active metasurface. **Figure 1a** depicts the target neural topology, emphasizing the dense, recurrent connectivity essential for spatiotemporal processing. **Figure 1b** shows its physical embodiment, where the network nodes are spatially defined by a programmable optical pump pattern. Each pump spot excites a localized microlaser operating at a quasi-BIC mode, which acts as a nonlinear photonic neuron. Critically, the BIC mode establishes strong, long-range radiative coupling between physically separated nodes [26], as shown in **Figure 1c**. Furthermore, the system natively incorporates intrinsic temporal memory. This is represented in **Figure 1a** by the "$x(t)$" labels, which correspond physically to the finite carrier lifetime of the gain medium and the photon lifetime between nodes (coupling-induced delay). These intrinsic time scales provide a built-in, analog recurrent feedback loop, obviating the need for any external delay lines or electronic memory [30]. Consequently, the relative timing and spatial arrangement of the optical pump dynamically reconfigure the effective synaptic weights, implementing a reconfigurable recurrent topology in hardware. Input signals encoded in the pump patterns and pulses sequences are thus nonlinearly projected into a high-dimensional state space shaped by collective lasing dynamics.

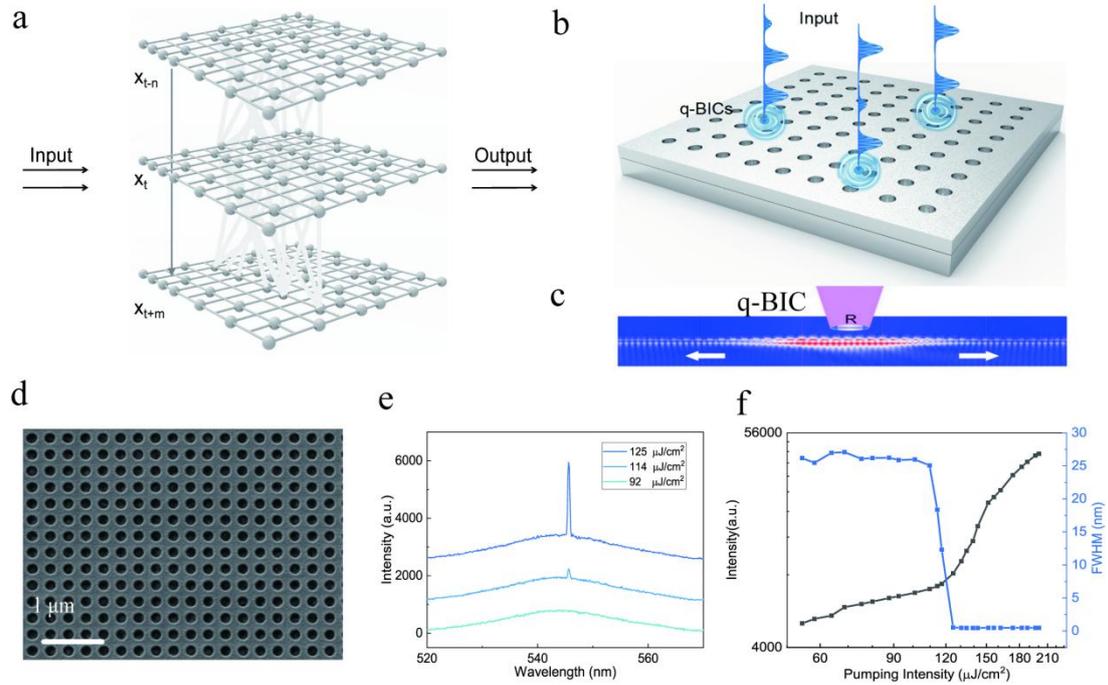

**Figure 1**. Photonic neuromorphic computing enabled by collective modes in coupled quasi-BICs. (**a**) Operating principle of neuromorphic computing enabled by a network of coupled nonlinear nodes, the layers denote the network states at different time steps.(**b**) Physical implementation of neuromorphic computing enabled by coupled quasi-BICs: femtosecond pumping at different time delays (indicated by the layer stacking) excites distinct collective supermodes in the BIC metasurface. (**c**) Schematic illustration of long-range coupling mediated by quasi-BIC mode across the metasurface. (**d**) SEM image of the fabricated BIC metasurface. (**e**) Emission spectra under increasing pump fluence. (**f**) Output intensity (blake line) and full width at half maximum (FWHM, blue line) as a function of pump intensity.

To physically implement the proposed neuromorphic architecture, we fabricated an active metasurface designed to operate at the symmetry-protected BIC mode[26]. **Figure 1d** shows the top-view SEM image of the sample, the device consists of a periodic array of air holes patterned into a 150-nm-thick ZEP520A resist layer on top of an intact 80-nm quasi-2D perovskite film ($N_2F_8$), covering an area of approximately 100 x 100 μm². Notably, we employed an etch-free fabrication process (Supplementary Note 1) to preserve the intrinsic optical and electronic quality of the perovskite gain medium, thereby minimizing structural perturbation [26,31]. The fabricated metasurface supports the targeted BIC modes, which are evident from its lasing characteristics under optical pumping. Under a femtosecond laser excitation, we observe a sharp transition from spontaneous emission to coherent lasing when increasing the pumping power. **Figure 1e** displays the evolution of the emission spectra. At low fluences, broad photoluminescence dominates the spectrum. Once the fluence reaches the lasing threshold ($P_{th} \approx 100$ uJ cm$^{-2}$), narrow lasing peaks emerge and rapidly dominate at higher excitation, accompanied by strong spectral narrowing. This threshold behavior is further corroborated by the clear "kink" in the input–output curve (**Figure 1f**). Additionally, the far-field emission exhibits a characteristic ring-shaped pattern (Supplementary Note 2), consistent with the radiation profile of a quasi-BIC mode and confirming the formation of radiatively coupled microlasers across the metasurface. Importantly, the input–output response provides an intrinsic, nonlinear transfer function that enables each microlaser to serve as a neuromorphic node with a built-in activation nonlinearity.

Crucially, the coupling between different quasi-BIC microlasers is strongly distance-dependent and highly sensitive to the resonators' relative positions and orientations within the lattice. For instance, coupling is significantly stronger along the Γ–X crystallographic direction than along Γ–M (Supplementary Note 3). This spatial anisotropy breaks the symmetry of the network connectivity, providing a directed-connectivity motif analogous to directed synapses in neural networks. For example, we configure three quasi-BIC resonantors into a triangle network (labeled "A" "B" "C" in **Figure 2a**), where the A–B pair is aligned along Γ–X, while A–C and B–C are aligned along Γ–M direction, resulting in intentionally asymmetric links. Theoretically, when the excitation is near the lasing threshold, the gain saturation is absent and the interaction between a set of coupled quasi-BIC microlasers can be modeled by the coupled-mode theory. When the coupling coefficients are all same between different resonators, the eigenvalues are degenerated. The resulting anisotropic coupling matrix in the BIC metasurface produces non-degenerate collective supermodes ($\psi_1$, $\psi_2$, $\psi_3$), i.e., the eigenstates of the network. As shown in **Figure 2b**, we calculated the eignenvaluse of the coupled system under eight input combinations ("000" - "111", "0" represent weak pumping/low gain and "1" represent strong pumping/high gain). The points with the same color in the figure represent the supermode eigenvalues when the input intensities at the three nodes are the same combination, where eigenmodes with positive imaginary parts correspond to net-gain (lasing-capable) states, while the real parts set the emission wavelengths. In figure 2b, we set the coupling coefficients between different resonantors to be asymmetric, yielding distinct modal gains and spectral positions (See Methods and Supplementary Note 4). Experimentally, we validate this by addressing each of the three nodes with binary optical inputs ("1" at $1.2P_{th}$ and "0" at $0.8P_{th}$). We tested multiple three-node geometries/coupling configurations. The configuration shown in **Figure 2c** is a representative case in which the eight input states ("000" to "111") yield clearly distinguishable output responses (other configurations can be seen in Supplementary Note 3). These combination of input patterns excite different weighted mixtures of the collective supermodes, leading to spatial-dependent nonlinear lasing states. **Figure 2c** and **2d** present the spectrally integrated intensity and the corresponding emission spectra, respectively, with near-field patterns shown alongside. It can be seen the experiment exhibit mode splitting and asymmetric intensity distributions that evolve systematically with the different inputs, consistent with selective activation among $\psi_1 - \psi_3$. The observed output variations, which are distinct for each input, signify an input-controlled redistribution of energy, highlighting the system's spatial processing capability.

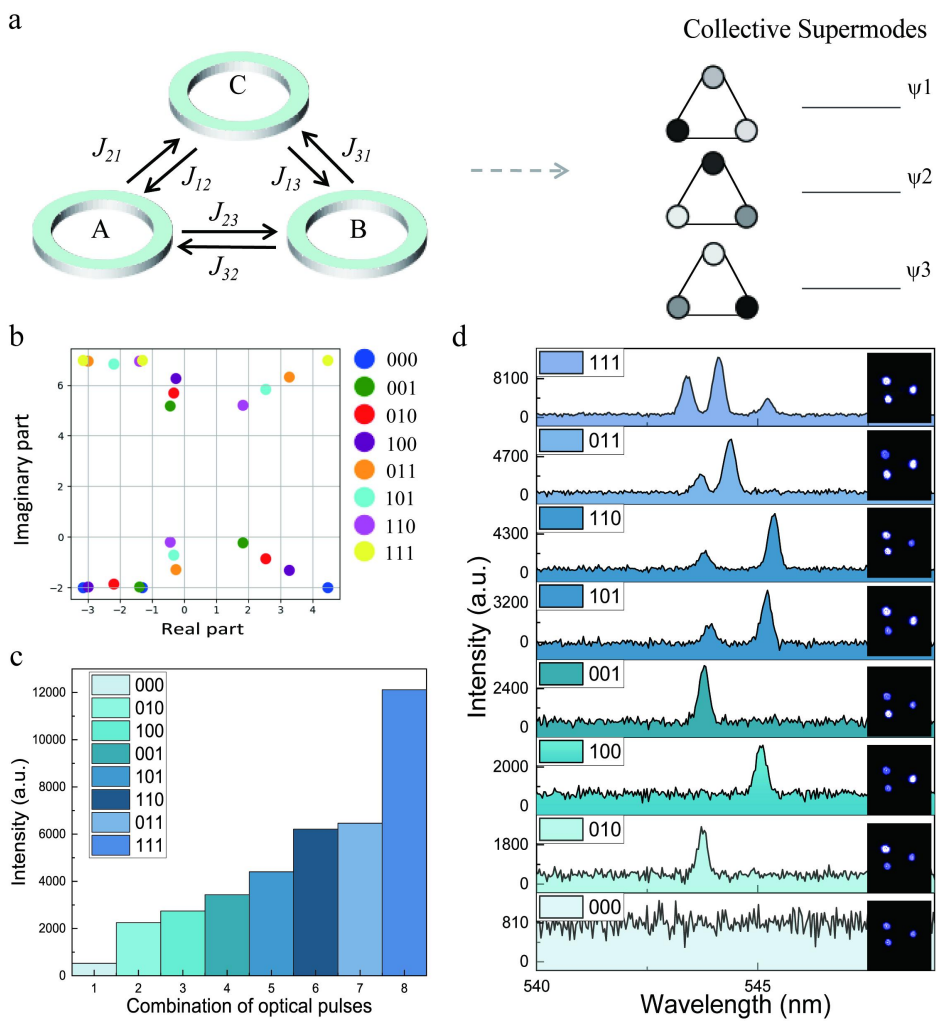

**Figure 2**. Spatially dependent responses of coupled quasi-BIC nonlinear optical neurons. (a) Schematic of three coupled quasi-BIC microlaser nodes under spatially asymmetric excitation, leading to an asymmetric coupling among the nodes. (b) Eigenvalues of the three-node coupled system under asymmetric coupling, the points with the same color in the figure represent the supermode eigenvalues when the input intensities at the three nodes are the same combination. (c) Spectrally integrated intensity near the lasing emission wavelength (540-550 nm) under eight different input conditions. (d) Emission spectra under eight different input conditions. On the right, corresponding near-field patterns are displayed, with each bright spot indicating pumping at $1.2P_{th}$ and each dark spot indicating pumping at $0.8P_{th}$.

Beyond the spatially anisotropic coupling, the network's collective response is also influenced by the temporal response of the nodes, making the effective interaction dependent on the relative timing between excitations [30]. To quantify this intrinsic temporal response, we employed a paired-pulse excitation scheme with a variable time delay $\Delta t$ and recorded the time-resolved emission using a streak camera, as illustrated in **Figure 3a**. This method allows us to quantify the system's memory through the paired-pulse facilitation (PPF) index [28], defined as PPF= $A_2/A_1$, where $A_1$ and $A_2$ represent the peak intensities of the transient emission signals in response to the first and second pulses, respectively.

We first investigate the response when both pump pulses excite the same spatial location. When an individual pulse exceeds the threshold, the device emits a delayed lasing pulse following a brief buildup period [30]. As shown in **Figure 3b**, when two above-threshold pulses (P=1.2 $P_{th}$, set at 0 ps and 50 ps, respectively) pump the sample, clear emission peaks are observed at t=55 ps and 120 ps in the streak camera's response. Since carrier relaxation and recombination processes in the perovskite gain medium occur over a timescale of several tens of picoseconds, the residual excited carriers generated by the first pulse enhance the gain of the second excitation, resulting in the second emission peak significantly higher than the first, revealing a graded facilitation effect. Additionally, when two individually sub-threshold pulses (P = 0.8 $P_{th}$) arrive at a fixed time delay, as shown in **Figure 3c**, the first pulse, being below threshold, does not produce a distinct lasing peak. But they can cooperatively exceed the lasing threshold, generating a transient lasing output only after the second pulse. **Figure 3d–e** plots the PPF index versus delay time ($\Delta t$) for above- and below-threshold conditions, respectively. In the above-threshold case, the PPF index follows an exponential decay, and a significant enhancement can still be observed at $\Delta t$=80 ps. This timescale corresponds directly to the carrier recombination lifetime of the perovskite gain medium. For two individually subthreshold pulses, $A_1$ is determined by referencing the peak position obtained from the above-threshold measurement under identical timing conditions. The dependence of the PPF index on $\Delta t$ is shown in Figure **3e**, which exhibits a decay trend similar to that in Figure **3d**. These decay times reveal a critical local memory window within a single photonic neuron and indicates that the system's temporal integration capability is limited by the carrier lifetime. Besides, the temporal response between spatially separated nodes is different. We found the nonlocal memory window is set by the persistence of the lasing pulse itself. (See supplementary note 3)

Beyond pairwise interactions, the system exhibits richer dynamics when processing longer pulses sequences. For 3-bit input sequences, we also set 8 possible signal encoding combinations, ranging from "000" to "111" ("1" at $1.2P_{th}$ and "0" at $0.8P_{th}$). As shown in Figures **3e-f**, when there is no delay between the pump pulses, the output intensity after pumping the sample with a set of pulses having the same total energy shows almost no difference. However, when a fixed time delay is introduced between the pulses, different sequences of preceding pulses lead to markedly different outputs due to the system's finite carrier lifetime and gain recovery dynamics. As shown in **Figure 3f** and **3g**. These output variations exhibit unique patterns for distinct input sequences, indicating the system's temporal processing capacity and adaptive response to varying input signals.

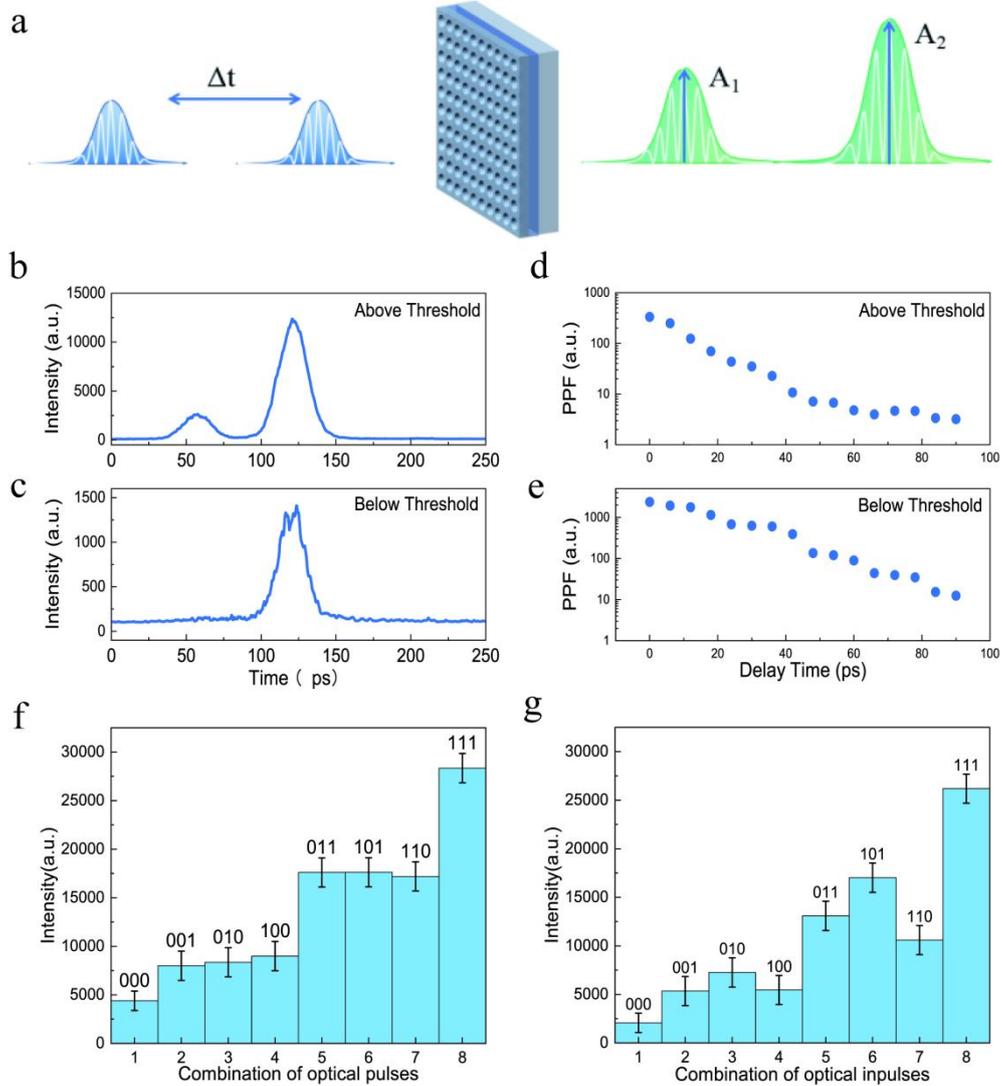

**Figure 3**. Temporal response in a BIC metasurface. (a) Schematic for time-resolved emission recorded using a streak camera. (b) Emission peaks in response to above-threshold excitation when two pulses excite the same sample spot. (c) Emission peaks for the same configuration in (b) under below-threshold excitation, with input pulses set at 0 ps and 50 ps, respectively. (d-e) PPF index plotted as a function of Δt, corresponding to (b-c). (f) Output intensities of the sample produced by the 3-bit binary optical pulses at Δt=0 ps. (g) Output intensities of the sample produced by the 3-bit binary optical pulses Δt=6.6 ps. The error bars indicate the intensity error caused by the instability of the pump laser.

**Neural Network Training and Testing Results:**

Having demonstrated the nonlinear spatiotemporal dynamics, we next evaluated the computational utility of the quasi-BICs network, we first implemented a reservoir computing (RC) framework and applied it to a static image classification task. The RC paradigm is ideally suited for the physical system, as it leverages the network's intrinsic, high-dimensional transients as a computational resource, requiring only the training of a simple linear readout layer [32-34]. Through the long-range, anisotropic coupling of the BIC metasurface, each distinct input pattern drove the network into a specific lasing state. The resulting emission—shaped by gain competition and the asymmetrical coupling matrix—encoded a high-dimensional projection of the input, which served as the reservoir state for subsequent linear classification. As outlined in **Figure 4a**, we selected a brain MRI tumor detection task using a publicly available dataset [35]. Each image was preprocessed into a standardized 30 × 30 binary matrix, and the image data was encoded spatially: consecutive triplets of pixel columns were grouped into 3-bit vectors (e.g., 100, 010...) to interface with the network. Physically, each vector was mapped to a corresponding spatial pump pattern projected onto the designed three quasi-BICs coupled system. Specifically, the output is a nonlinear function governed by the dominant collective supermode under a given pump configuration. Reservoir states are then accessed through two optical readouts: the full emission spectrum and the spectrally integrated intensity, as shown in **Figure 2**.

    **Figure 4b–d** present the results using the integrated intensity as the reservoir states (See Supplementary note 4 for setails of neural network training and testing). This single-scalar readout achieves a test accuracy of 90.2%, with stable training convergence illustrated in **Figure 4b**. And the performance is reflected in the clear class separation visible in the confusion matrix (**Figure 4c**) and is quantitatively supported by a receiver operating characteristic (ROC) curve with an area under the curve (AUC) of 0.87 (**Figure 4d**). Figure **4e–g** presented the results when the full optical spectrum is used as the reservoir state. This scheme yields a higher test accuracy of 92.16% (**Figure 4e**), with tighter clustering in the confusion matrix (**Figure 4f**) and a stronger ROC curve exhibiting an AUC of 0.89 (**Figure 4g**). To validate the generality of this approach, we repeated the MRI classification task for multiple distinct three-node geometries (obtuse, equilateral, right, and acute triangles). Across all tested geometries, the spectral readout consistently achieved high accuracy (91.3% ± 1.2%), confirming that the quasi-BIC reservoir provides robust nonlinear mapping regardless of specific node arrangement. These results demonstrate that while topology can be optimized for maximum performance, the system's computational capability is not critically sensitive to exact node placement, offering flexibility for practical implementation.

    We note there is a critical system-level trade-off exists between the feature richness of the spectral readout and its acquisition speed. Although spectrally resolved measurement provides higher dimensionality, it is inherently slower. Therefore, for the subsequent dynamic task—which demands maximal temporal throughput—we employed the faster integrated-intensity readout to leverage the system's intrinsic picosecond-scale processing without introducing a bottleneck. Notably, the platform demonstrated here is inherently scalable and programmable across multiple physical dimensions. Employing digital micromirror devices or spatial light modulators would

enable independent and parallel pumping of a larger number of discrete quasi-BIC nodes or even continuous patterns (Supplementary Note 4).

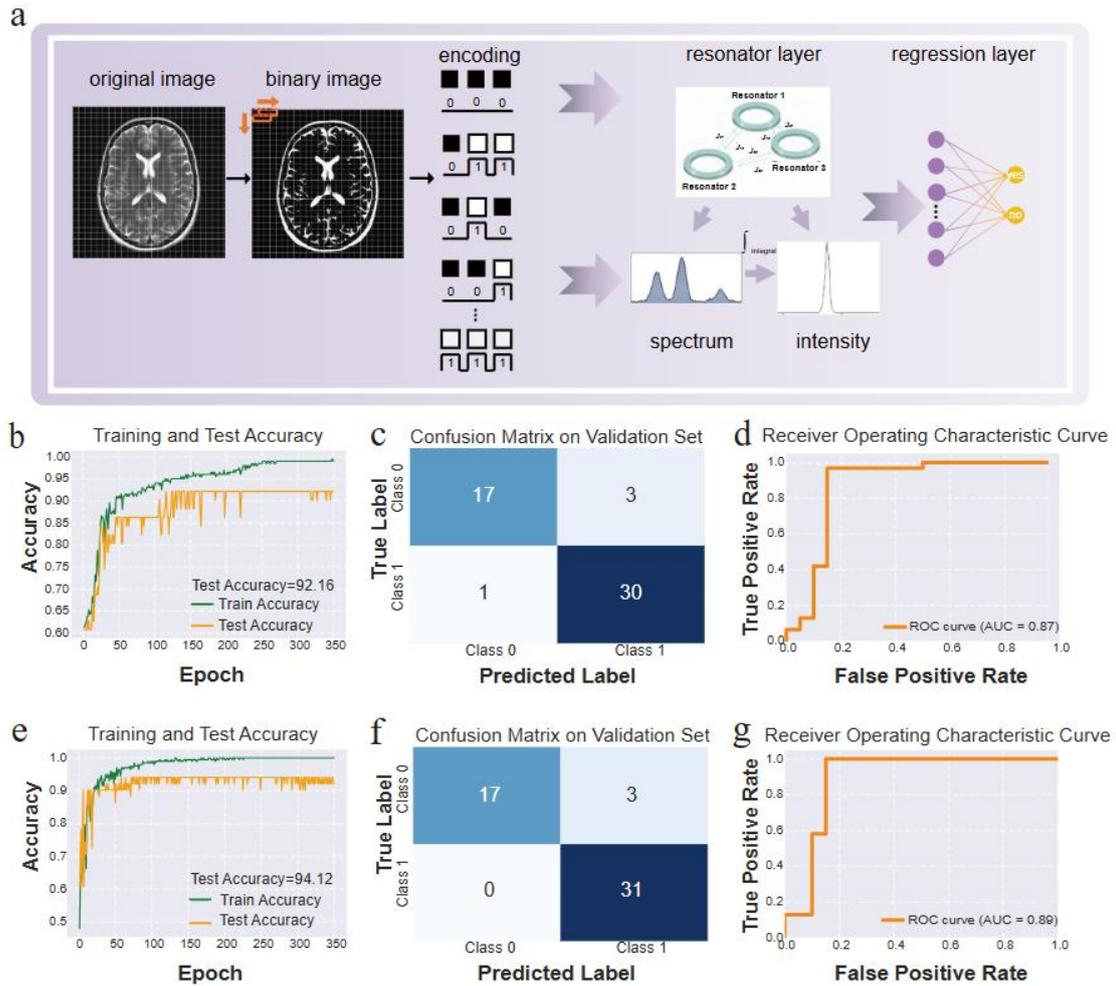

**Figure 4**. All-optical reservoir computing based on coupled quasi-BIC microlasers for MRI image classification. (a) Schematic of the photonic reservoir framework. Binary-encoded brain images are converted into spatial pump patterns projected onto a three-node quasi-BICs network, whose multimode lasing responses serve as reservoir states for linear-readout classification. Two readout schemes were evaluated: integrated intensity (b–d) and optical spectrum (e–g). (b,e) Training and validation accuracy curves showing stable convergence. (c,f) Confusion matrices on the validation set demonstrate reliable discrimination between tumor and non-tumor samples. (d,g) Receiver-operating-characteristic (ROC) curves further confirm strong classification capability, with the integrated-intensity and spectral readouts achieving overall accuracies of 90.20 % and 92.16 %, corresponding AUC values of 0.87 and 0.89, respectively.

Having demonstrated the capability for static image classification, we now exploit the full spatiotemporal processing capacity the BIC reservoir. By programming the relative time delays among spatially encoded pump pulses, we can harness these transient dynamics to physically encode both spatial patterns and their temporal evolution in a single optical transformation. We validate this concept using the NTU RGB+D skeleton-based action dataset [36], and we selected ten action categories from the dataset to evaluate the network's recognition performance. As outlined in **Figure 5a**, the dataset provides 3D skeletal coordinates for 25 body joints per frame. We reconstructed these coordinate sequences into binary video streams, where joint positions were assigned a value of "1" and all other regions "0", yielding a spatiotemporal binary representation with a resolution of 20 × 12, as shown in **Figure 5b**. To adapt the data for hardware processing, we standardized the variable-length videos in the temporal domain: first, by uniformly sampling at a fixed interval $\Delta t$ to obtain a fixed length of 30 frames per sample; subsequently, we applied a three-frame differencing technique [37,38] to enhance motion saliency by highlighting dynamic changes between consecutive frames.

The core of data encoding lies in pre-measured, device-specific nonlinear transfer functions in spatial and temporal domain. For a given video sequence, the spatial pattern is then transformed by the spatial kernel, and the temporal evolution across frames is convolved with the temporal kernel, as shown in **Figure 5c and 5d.** To quantify the contribution of spatial and temporal nonlinearities, we performed a controlled study using the action dataset. We construct four processing scenarios for action recognition: Scenario A (Linear Baseline): Utilizes only the linear projection of raw binary pixels, disregarding all spatiotemporal correlations. Scenario B (Spatial-only Nonlinear): Employs only the aforementioned spatial nonlinear encoding with frame-independent processing. Scenario C (Temporal-only Nonlinear): Employs only the temporal nonlinear encoding with independent processing for each spatial point, ignoring spatial correlations. Scenario D (Spatiotemporal Nonlinear): Simultaneously enables both spatial and temporal nonlinear encoding, fusing $X_s$ and $X_t$ for unified processing, as shown in figure **5f**. The performance of different scenarios clearly delineate the contributions of spatial-temporal jointly encoding. In Scenario A, treating each video frame as an independent spatial pattern without exploiting temporal and spatial correlations. This baseline achieves a classification accuracy of 81.49 % (**Figure 5e**). When we take into account the nonlinear mapping in the temporal or spatial domain, the recognition accuracy of the neural network improves stochastically (See details in Supplementary note-4). Especially, in Scenario D (spatiotemporal encoding), video sequences are encoded as spatiotemporal pump patterns that engage both the network's spatial coupling and its temporal memory. This integrated approach raises the recognition accuracy to 85.36 %, as shown in **Figure 5f**, representing a +3.87 % absolute improvement. This enhancement stems directly from the physical network's native ability to process correlated information across space and time. Moreover, the loss function of the BIC resonant network converges notably faster than that of the conventional network, indicating that the physical-layer preprocessing significantly optimizes the distribution of the feature space and reduces the computational burden of downstream training. This combination of ultrafast optical feature extraction and low training cost fully demonstrates the great potential of optical BIC networks in high-performance, real-time visual processing tasks.

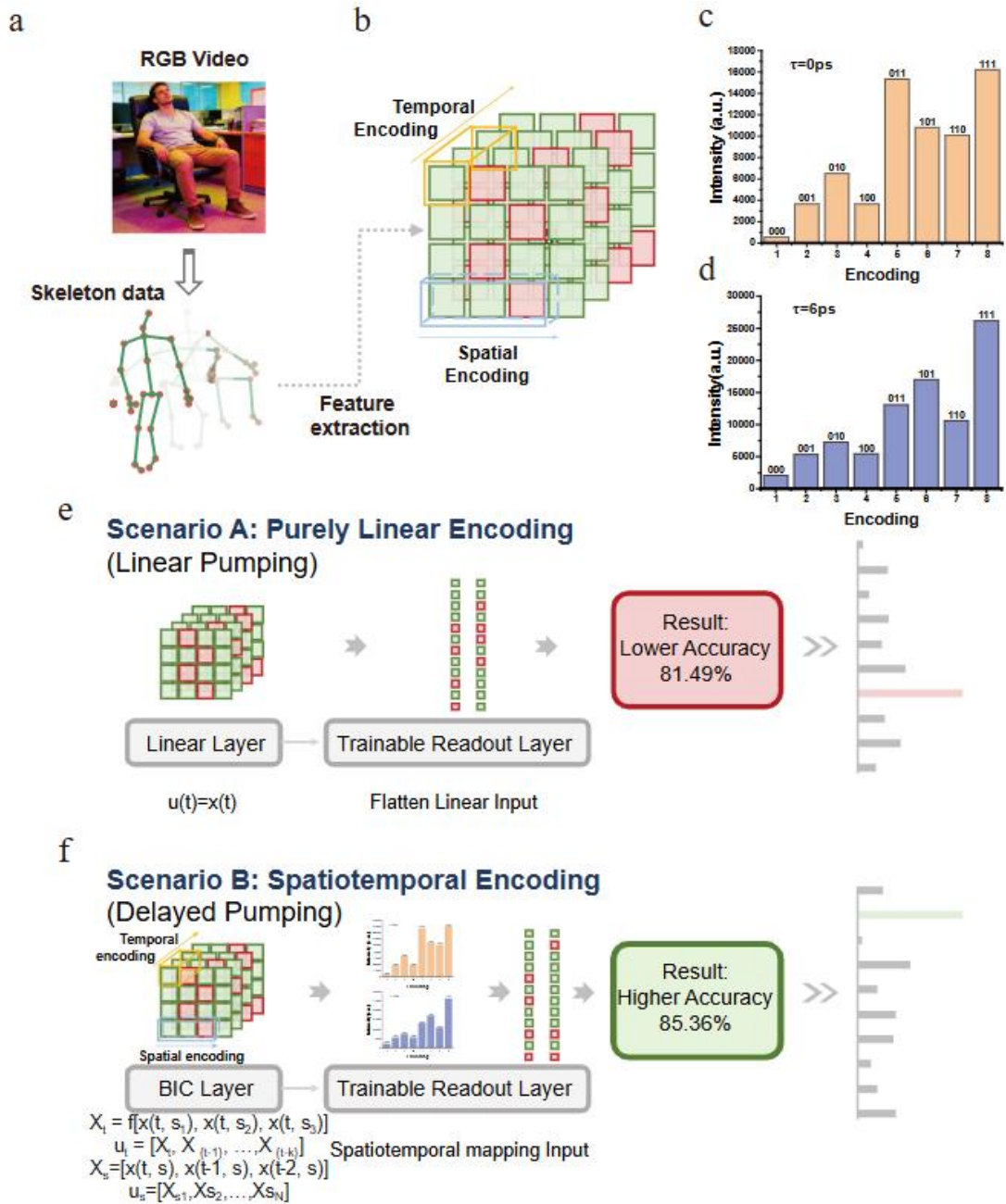

**Figure 5**. Spatiotemporal encoding and action recognition with the coupled quasi-BICs photonic reservoir. (a) Workflow of skeleton-based human action recognition using the designed network. (b) Preprocessing pipeline: skeleton coordinates are converted to binary video frames, temporally down-sampled to k frames with interval Δt, and enhanced via three-frame differencing. (c) Experimentally measured integrated intensity responses when three pump pulses, with all eight possible temporal on/off patterns (000–111), are delivered to the same spatial location. (d) Integrated intensity responses for all eight spatial pump patterns (000–111) applied simultaneously to three different locations. (e) Classification accuracy for purely linear encoding. (f) Classification accuracy for spatiotemporal encoding, which jointly utilizes both spatial patterns and temporal patterns through the network's nonlinear mapping.

**Discussion and Conclusion**

Above we demonstrated a compact and efficient photonic neuromorphic computing architecture based on the BIC photonic reservoir. Using a minimal yet physically complete three-node system, we validate the core computational primitives of nonlinear activation, reconfigurable long-range coupling, and analog temporal memory. Through tailored input-encoding and data-preprocessing strategies, this elementary unit was successfully applied to benchmark tasks such as image classification and video action recognition. At the same time, we recognize that the current work has room for improvement in terms of node scale, multi-timescale processing capability, system cascadability, and energy consumption. However, these limitations are primarily technical rather than fundamental. For instance, the physical mechanisms revealed here can be transferred to other mature platforms, such as III-V semiconductors [17] and polariton-based networks [39,40]. These platforms are expected to support continuous-wave operation and achieve lower thresholds. Regarding temporal processing, the limited carrier lifetime of the current perovskite gain medium restricts the complexity of temporal dependencies that can be captured within a single processing window. Future integration of gain materials with longer excitation lifetimes, such as rare-earth-ion-doped dielectrics [41]. Furthermore, the etch-free fabrication process offers significant engineering advantages, and the periodic translational invariance of the structure effectively avoids complex optical alignment challenges. Subsequent development of electrically pumped BIC microlaser arrays, combined with an optical injection-locking architecture [21,24], could enable signal propagation and nonlinear transformation across multiple stages at the same optical wavelength, laying the foundation for building deep cascaded, trainable all-optical information-processing systems.

In summary, this study demonstrates the BIC metasurface can function as an efficient physical reservoir, providing a novel pathway toward all-optical neuromorphic intelligence. This framework unifies spectral richness, nonlinear spatial coupling, and built-in temporal dynamics within a single metasurface, enabling direct hardware-level processing of spatiotemporal information. Such systems are not merely hardware accelerators for fixed algorithms but are promising as ultrafast sensing–computing fusion engines, capable of real-time processing of complex data streams in vision, lidar, and even quantum information domains, ultimately delivering disruptive hardware solutions for edge AI and real-time decision-making.

**Methods:**

**Sample Preparation.** We designed the BIC unit cell using commercial software (COMSOL Multiphysics). Periodic boundary conditions were applied in the x- and y-directions to mimic an infinitely large structure, while perfectly matched layers in the z-direction absorbed outgoing waves. Optical constants for the active layer and top polymer were obtained through ellipsometry measurements of Quasi-2D perovskite and ZEP520A resist. The refractive index of the glass substrate was set to 1.45. The BIC resonant metasurface was fabricated on a 13 nm ITO-coated glass substrate through a combination of spin-coating and electron beam lithography. We prepared the precursor solution of $N_2F_8$ ($(NMA)_2FA_{n-1}PbBr_{3n+1}$, n=8) by dissolving a 25% molar ratio of 1-naphthylmethylamine bromide (NMABr) in a mixture of $HC(NH_2)_2Br$ (FABr) and $PbBr_2$ (1:1 ratio) in DMF at 0.4 M, stirred at 60 °C for 12 hours. The experimental details have been summarized in Supplementary Note-1.

**Optical Characterization.** The sample was optically excited by a frequency-doubled Ti:Sapphire laser (Spectra-Physics, repetition rate 1 kHz, pulse width 100 fs). Its output was frequency-doubled through a β-barium borate (BBO) nonlinear crystal to generate 400-nm excitation pulses. The emission spectrum was recorded by fiber-coupled spectrometer (Ocean Insight FX-XSR), and the time-resolved emission signals were recorded by a streak camera (XIOPM 5200). The details of the optical setup have been summarized in Supplementary Note-2.

**Theory analysis.** Considering that all pumping powers are near the threshold, gain saturation is negligible. Consequently, the interactions among a set of coupled quasi-BIC microlasers can be effectively described using coupled mode theory. For multiple parallel inputs, we extend this framework to $n$ quasi-BIC cavities. Denote $G=\text{diag}(g_1, g_2, g_n)$ represent the net gain of each resnonatro and $J$ the symmetric coupling matrix set by the pump pattern. $|\psi\rangle=(a_1,a_2,...,a_n)^T$ represents the states vector, where $a_i$ menas the complex amplitude of the $i$-th resonator. The coupled-mode equation is given by:

$$i\dot{\psi}=(G+J)\psi \qquad (1)$$

with modal decomposition:

$$(G+J)V_k=\lambda_k V_k \qquad (2)$$

and the time evolution represented as:

$$\psi(t)=\sum_k C_k e^{\lambda_k t} V_k \qquad (3)$$

Thus, a given pump pattern (encoding) sets $G$, which, together with the fixed $J$, defines a spectrum $\{\lambda_k\}$ and supermodes $\{V_k\}$. The system lases in the supermodes with the largest real part of $\lambda_k$. Distinct spatial pump profiles remove degeneracies, enabling non-degenerate multimode emissions. Both the intensity and spatial distribution encode the input, while the collective emissions facilitate a high-dimensional nonlinear mapping from the pump space to the optical output. This output can then be analyzed spectrally or in terms of intensity and processed electronically to achieve the target task.

**Data Processing.** To ensure a fair comparison between different encoding scenarios (linear, spatial‐ nonlinear, and spatiotemporal‐ nonlinear), we strictly controlled the input dimension of the final linear regression layer. For all scenarios, the processed feature vector was projected onto a fixed- dimensional space of size D = 512 (or another number) before classification. Specifically:For the Linear Baseline (Scenario A), the flattened video frames were used as raw features. For the Spatial- Nonlinear scenario (Scenario B), each frame was transformed by the pre- characterized spatial BIC kernel, and frame- level features were concatenated. For the Spatiotemporal- Nonlinear scenario (Scenario C), features from both the spatial and temporal kernels were combined. In all cases, a principal component analysis (PCA) transformation, fitted once on the training set of the Linear Baseline scenario, was applied to reduce/expand the feature vectors to the common dimension D. This guarantees that any performance difference originates solely from the informational content and nonlinear transformation of the features, not from the dimensionality or capacity of the classifier

**Acknowledgement**

The authors acknowledge support by National Key Research and Development Program of China (Grant Nos. 2024YFB2809200, 2021YFA1400802 and 2022YFA1404700), National Natural Science Foundation of China (Grant Nos. 6233000076, 12334016, 11934012, 12025402, 62125501, 12261131500, 92250302 and 62305084), Guangdong Basic and Applied Basic Research Foundation (2023A1515011746, 2024B1515020060), Shenzhen Fundamental Research Projects (Grant Nos. GXWD 2022081714551), Guangdong Provincial Quantum Science Strategic Initiative(GDZX2406002), Shenzhen Science and Technology Program (Grant Nos. JCYJ20230807094401004), Shenzhen Fundamental research project (JCYJ20241202123719025, JCYJ20210324120402006, JCYJ20200109112805990), the Fundamental Research Funds for the Central Universities (Grant No. HIT.OCEF.2024020, 2022FRFK01013).

**Data availability**

The data that support the findings in this study is available from the corresponding authors upon reasonable request.

**Competing interests**

The authors declare no competing interests.

**Author contributions**

Q.S.,C.H. conceived the idea and supervised the research. J.F., H.T., Z.X.,Y.C., did the simulation. H.T., Z.X., X.J., R.J. fabricated the samples. Z.X. R.J,. H.T., T.K. and G.Q. performed the experimental measurements. C.H. prepared the manuscript with contributions and discussions from all authors.